\documentstyle[12pt]{article}
\begin{document}
\setlength{\baselineskip}{15pt}
\begin{titlepage}

\vspace{5mm}

\begin{center}
 Fractal dimension and degree of order in sequential deposition of 
mixture
\end{center}

\vskip 8mm

\begin{center}

M K Hassan
\end{center}
\begin{center}
Department of Physics, Brunel University, Uxbridge, Middlesex UB8
  3PH, United Kingdom
\end{center}

\vskip 8mm

\begin{center}
{\bf Abstract}
\end{center}

\noindent
We present a number models describing the sequential deposition of a mixture 
of  particles whose size distribution is determined by the power-law 
$p(x) \sim \alpha x^{\alpha-1}$, $x\leq l$ . We explicitly obtain the  
scaling 
function in the case of random sequential adsorption (RSA) and show that 
the pattern created in the long time limit becomes scale invariant. This
pattern can be described by an unique exponent,
the fractal dimension. In addition, we introduce an external tuning 
parameter 
$\beta$ to describe the correlated sequential deposition of a mixture of 
particles 
where the degree of correlation is determined by $\beta$, 
while $\beta=0$ corresponds to random sequential deposition of mixture. 
We  show that the fractal dimension of the resulting 
pattern increases as $\beta$ increases and reaches  a constant 
non-zero value in 
the limit $\beta \rightarrow \infty$  when the pattern becomes 
perfectly ordered or non-random fractals.   

\vspace{25mm}


\noindent
Permanent address: \\
\noindent
Department of Physics, Shahjalal Science and Technology
University, Sylhet, Bangladesh

\noindent
E-mail addresses: Kamrul.Hassan@brunel.ac.uk

\end{titlepage}

\section{Introduction}

The formation of stochastic fractals is an active field of research
both theoretically and experimentally. Yet the mechanism by which nature 
creates fractals and the relationship between the degree of order and 
the fractal dimension is poorly understood.    
The history of describing natural objects by geometry is as old as the  
science itself. However, traditionally Euclidean lines, squares, 
rectangles, circles, spheres etc. have been  the basis of our intuitive
understanding of the geometry of almost all  natural objects. But,
nature is not restricted to Euclidean space. Instead, most of the natural 
objects we see around us are  so complex in shape that conventional 
Euclidean space is not sufficient to describe them. It appears to be 
essential to invoke the concept of fractal geometry to characterize such 
complex objects quantitatively. It further 
enables us to search for  symmetry and order even in  disordered,
 complex systems \cite{kn:has91,kn:sapoval}. The importance of the 
discovery of fractals can hardly be exaggerated. Yet, there is no neat 
and complete 
definition of a fractal. Instead one associates a fractal 
with a shape made of 
parts similar to the whole in some way. It is typically quantified  by a 
non-integer exponent called the fractal dimension that can uniquely 
characterize the structure. This 
definition immediately confirms the existence of scale invariance, that is,
objects look the same on different scales of observation. 
To understand fractals, their physical origin and 
how they appear in nature we need to be able to model them theoretically. 
This forms part of our motivation of the present work.

The simplest way to construct a fractal is to deterministically repeat a 
given operation.  The construction of a classical Cantor 
set is a simple text book example of such a fractal. It is created by 
fragmenting a line into $n$ equal pieces and removing $n-m$ of the parts 
created and repeating the process with the $m$ remaining pieces 
\cite{kn:has91}. This process is repeated 
{\it ad infinitum}. However, this construction is rather too artificial 
as it differs  in two ways from the fractals that occur in nature. It 
does not have any kinetics and it does not have any  randomness.

In this work, we introduce a stochastic process allowing a number of 
intrinsic tuning parameters which 
may be considered as a natural kinetic counterpart of the classical
Cantor construction and which is a potential candidate in order to 
understand the essential governing rule of  creating complex objects. These 
intrinsic tuning parameters are used to determine the degree of randomness  
and the rate at which a given operation is repeated to 
create a fractal. The interval  are chosen for breaking  
stochastically and once an interval is chosen the cuts are placed 
randomly on the interval while the degree of randomness is determined by  
the precise choice of deposition kernels. Thus starting 
with an infinitely long interval, in the long time limit what remains 
in the long time limit are an 
infinite number of points scattered over the intervals. The properties of 
these  points create a set that appears to be  statistically self-similar 
and is characterized by a fractal dimension. 

The construction of stochastic fractals we consider is not at all 
pedagogical. One immediate and potential application of the stochastic 
fractal is  the sequential deposition of a mixture of particles with 
continuous  distribution of sizes. 
 However, the model we consider  in this work mimics the configuration when 
objects once inserted 
are clamped  in their spatial positions for which non-equilibrium 
configurations are generated. 
 The  RSA processes
have been found to describe many experimental  systems, namely, the
adhesion of proteins and colloidal particles to  uniform surfaces
\cite{kn:has60,kn:has61}, the reaction of various polymer chain systems  
such as methyl vinyl ketone \cite{kn:has62} and many fields  in chemistry  
and physics. Although the process is 
conceptually simple, understanding its kinetics analytically is  a 
challenging problem (see an
excellent review article \cite{kn:has92}). The deposition of particles of 
definite sizes in one dimension  has been solved exactly and analytically 
in both continuous and discrete cases. The continuous version of this 
model is known as the random car parking problem.  
 Recently, the deposition of a mixture of particles, 
such as deposition of two particles of different sizes has been 
considered  \cite{kn:bonnier,kn:epstein,kn:rodgers}. From 
the experimental  point of view, the deposition kinetics of well-defined 
mixtures of different types of particles is of great importance.  It is 
well known that if  objects 
are of definite size and once deposited are clamped in their positions, 
non-equilibrium configurations  are created with a strong non-ergodic 
and non-Markovian nature \cite{kn:ramsden}. In this case, 
the resulting system  does not produce a 
scale invariant pattern. Instead, it reaches  a jamming limit when it is 
impossible to place further objects without overlapping. Hence a unique 
number (the jamming limit or the coverage) is sufficient to characterize the 
resulting pattern created in the long time. If a mixture 
contains a  continuous  distributions of sizes and is deposited 
sequentially, such that once deposited it is clamped to its position, then 
it is equivalent to our stochastic kinetic Cantor construction 
\cite{kn:krapivsky}. Hence the 
deposition of a mixture of particles will create a scale invariant pattern 
that does not reach the jamming limit since there always exists  a 
particle if there is an uncovered space during the process.

Interesting questions arises from the present work: (i) what is the 
role of the fractal dimension during pattern formation? (ii) Is there 
any relation between fractal dimension and degree of order? (iii) what 
are the relevant parameters to tune the degree of order and what are 
their physical meaning. The present work is an attempt to answer these 
questions. In  section one, we present the general equation to describe  
sequential deposition of an arbitrary number of particles. In 
section two, we present model I describing  random sequential deposition 
of a mixture of particles at a rate determine by the size of the 
empty space and exponent $\gamma$. We obtained explicit 
scaling solution and shown the resulting pattern in the long time 
limit is scale invariant both in time and space. In section three, we 
 consider correlated sequential deposition of a mixture of particles 
(model II). We 
further  obtain the fractal dimension of the resulting pattern for 
model II and 
compared with the fractal dimension for model I. Finally, we present 
a generalised version of both model to describe correlated sequential 
deposition of an arbitrary number of particles where, the degree of 
correlation  is controlled by a parameter $\beta$ and obtained fractal 
dimension as  a function of $\beta$. In the final section, we 
summarise 
the results discuss various point in order to reach a conclusion.

\vspace{5mm}

\section{Sequential deposition of arbitrary number of particles}

The connection between the $1$-d model of car-parking and the fragmentation 
processes was  first emphasized by Ziff \cite{kn:ziff} . The 
fragmentation process  can be thought of as the deposition of points on a 
line whose position  depends on the kinetic rule defined by the choice of 
kernels. Let $\psi(x,t)$ be the concentration of empty or 
uncovered intervals of length  $x$ at time $t$. Then the rate equation  
for this concentration obeys the integro-differential equation  
\begin{eqnarray}
{{\partial\psi(x,t)}\over{\partial t}} & = & -\psi(x,t)\int\prod_{j=1}^{n-m}p
(x_j)F(x_{n-m+1},..,x_n|x_1,..,x_{n-m})\delta(x-\sum_{i=1}^nx_i)\prod_{i=1}^n
dx_i + m  \nonumber \\ &  & \int 
\psi(y,t)F(x,x_{n-m+1},..,x_{n-1}|x_1,..,x_{n-m})\delta(y-x-
\sum_{i=1}^{n-1}x_i)\prod_{i=1}^{n-1}dx_idy\prod_{j=1}^{n-m}p(x_j) 
\nonumber \\ 
\end{eqnarray}
Here,  $n=3,4,...,$ and $m=2,3,...,$etc. so that $(n-m)=1,2,3,...,$etc is 
the number of particles which are 
deposited at each time step. Notice that the $m$ value in particular put a 
strong constraint on the depositing 
particles. That is, if $(n-m)=i$  and $m<(i+1)$ then the precise $m$ will 
determine how many of the depositing particles must 
deposit next each other so that they produce only $m$ new empty spaces.   
The term $p(x_i)$  determine the size of the depositing 
particles at each time step and $F(x_{n-m+1},..,x_n|x_1,..,x_{n-m})$ 
determine the rate and the rules with which 
$x_1,..,x_{n-m}$ particles are to be deposited to create $m$ new empty 
spaces at each time step. The first term on the right represents 
the destruction of spaces of size $x$ and the second term 
represents their creation from bigger spaces. Equation (1) can also 
describe the process of breaking an interval into $n$ pieces and throwing 
away $(n-m)$ pieces  to create a stochastic fractal. 

\section{Model I}

We first choose to consider the deposition of one 
particle at each time step ($m=2$), when the size of the depositing 
particles follows the power law form
\begin{equation}
p(x)=\alpha x^{\alpha-1} \hspace{5mm} {\rm for} \hspace{5mm} x\leq l.
\end{equation}
This  form of the parking distribution 
implies the deposition of
a mixture of particles with a continuous distribution 
of sizes. We further
choose  the rate with which particles are deposited 
to be
\begin{equation}
F(x_2,x-x_1-x_2|x_1) = x^\gamma.
\end{equation}
This choice of kernel implies that any point in the empty spaces are equally 
likely 
to be chosen for deposition by a particle whose size is determined by the 
choice of $p(x)$> However,   the empty space where the particle is 
deposited is determined by the exponent $\gamma$.  The rate equation
then becomes 
\begin{equation} 
{{\partial\psi(x,t)}\over{\partial t}}=
-{{x^{\alpha+\gamma+1}}\over{\alpha+1}}\psi(x,t)+2\int_x^l
\psi(y,t)y^\gamma (y-x)^\alpha dy \hspace{4mm} {\rm for} \hspace{4mm} 
x\leq l. 
\end{equation}
Notice that setting $\alpha =0$  describes the random deposition of
$zero$ sized particles on a line at a rate $x^{\gamma+1}$ i.e. 
the fragmentation process \cite{kn:has3}.
It is well known that  in one dimension the random  parking of cars of
length `1'  is highly non-ergodic in the sense that the whole space is not 
visited by the depositing particles. However, in  the case of  the random 
deposition of a mixture of particles with a
power law form of distribution sizes, the system retains an ergodic 
nature. In fact, the size of the particles to be deposited depends 
intrinsically on the size of the available empty spaces. Consequently, 
the size of the deposited particles becomes point like in the long time 
limit and there is always an available particle to deposit
whatever the size of the empty space in the system.  The second term on 
the right hand side of Eq. (4)  reveals that as $\alpha$ increases the 
non-local character becomes more and more prominent, so that it becomes 
increasingly difficult to solve the equation. Notice that the exponent 
$\gamma$ does not play any role in determining the degree of non-locality. 

We define the moment of the empty size distribution function as
\begin{equation}
M(q,t) =\int_0^l\psi(x,t)x^qdx
\end{equation}
We now multiply on both
sides of the rate equation by $x^q$  and integrate over $x$ thus
obtaining  a rate equation for the moments 
\begin{equation}
{{dM(q,t)}\over{dt}}=\Big (2{{\Gamma(q+1)\Gamma(\alpha+1)}\over{\Gamma
(q+\alpha+2)}}-{{1}\over{\alpha+1}}\Big ) M(q+\alpha+\gamma+1,t). 
\end{equation}
This equation can be solved to find the solution for the $qth$ moment,
the solution taking the form of a generalised hypergeometric
function with $(\alpha+1)$ numerator and denominator parameters.  In
order to understand the various physical aspects of the
problem within the simplest possible  way, we  first consider the case when
$\alpha=1$. In this case the rate equation  becomes

\begin{equation}
{{\partial\psi(x,t)}\over{\partial t}}=
-{{x^{\alpha+\gamma+1}}\over{\alpha+1}}\psi(x,t)+2\int_x^l
\psi(y,t)y^\gamma (y-x)^\alpha dy,
\end{equation}

This class of models includes the one that is considered in 
\cite{kn:krapivsky,kn:has94} when  $\gamma=0$.  Iterating the above
equation to get all the  higher derivatives and substituting in the
Taylor series expansion of $M(q,t)$ about $t=0$ yield
\begin{equation}
M(q,t)=l^q \
_2F_2\Big ({{q-a}\over{\gamma+2}},{{q+a+3}\over{\gamma+2}};{{q+1}\over
{\gamma+2}},{{q +2}\over{\gamma+2}}; -tl^{\gamma+2}\Big ),
\end{equation}
where, $ _2F_2(a,b;c,d;x)$ is the generalised hypergeometric function
\cite{kn:has86,kn:has93} and
$a=(-3+\sqrt{17})/2=0.5615288$. The asymptotic expansion of the 
generalised hypergeometriv function immediately reveals that the moments 
show the following power-law behaviour
\begin{equation}
M(q,t) \sim t^{{{q-a}\over{\gamma+2}}}.
\end{equation}
Notice that the exponent of the asymptotic expression for the moment is 
linear in $q$ which reveals the existence of simple scaling. 
 
\subsection{Explicit scaling solution}

In this section we attempt to show that in the long time limit the empty 
size distribution shows power-law behaviour. Linear power-law behaviour 
of the moments reveals that the system reaches to scaling or self-similar 
behaviour does not depend on 
the initial conditions for which one can invoke universality.  However, a 
closer look at the rate 
equation further reveals that only one of the two parameters has 
independent dimension i.e. $x$ and $t^z$ have the same dimension where 
the exponent $z$ is yet to be determined. 
That is, the dimension of $\psi$ must be expressible in terms of the 
independent parameter $t$ or $x$ alone. We can therefore define the 
dimensionless quantities as 
\begin{equation}
\xi= xt^{-{{1}\over{z}}} \hspace{3mm} {\rm or} \hspace{3mm} \zeta = tx^z
\end{equation}
and 
\begin{equation}
\phi(xt^{{1}\over{z}}) = {{\psi(x,t)}\over{t^{{\nu}\over{z}}}}, \hspace{3mm} 
\Phi(tx^z)= {{\psi(x,t)}\over{x^{-\theta}}}. 
\end{equation}
If scaling theory is obeyed a plot of 
${{\psi(x,t)}\over{t^{{\nu}\over{z}}}}$ against $\xi$ or a plot of 
${{\psi(x,t)}\over{x^{-\theta}}}$ against $\zeta$ should fall on a 
single curve for any initial distribution. This reveal also that 
 a self-similar solution in time and space exists. However, we find 
it convenient to consider the following scaling {\it ansatz} 
\begin{equation}
\psi(x,t) =x^{-\theta}\Phi(tx^z)
\end{equation}
for which the pattern that develops in the long time limit is 
self-similar in space.  
The exponent $\theta$ can be obtained from the rate equation for the 
moments using the condition for  moment is time independent to give 
$\theta=1+a$. We 
now substitute  this {\it ansatz} into the rate equation to obtain 
\begin{equation} 
t^{{{\gamma+2-z}\over{z}}}= 
{{-{{1}\over{2}}\zeta^{{{\gamma+2-\theta}\over{z}}}\Phi(\zeta)-
{{2}\over{z}}\zeta^{1/z}\int_\zeta^\infty
\eta^{{{\gamma+1-z-\theta}\over{z}}}\Phi(\eta)d\eta 
+{{2}\over{z}}\int_\zeta^\infty\eta^{{{\gamma+2-z-\theta}\over{z}}}
\Phi(\eta)d\eta}\over{\zeta^{{{z-2}\over{z}}}\Phi^\prime(\zeta)}}.
\end{equation}
Demanding the scaling solution to exist we find that $z=\gamma+2$. The  
equation that we need to solve to find the scaling
solution is
\begin{equation}
\zeta^{{{\gamma+2-\theta}\over{\gamma+2}}}\Phi^\prime(\zeta)+ 
{{1}\over{2}}\zeta^{{{\gamma+2 -\theta}\over{\gamma+2}}}\Phi(\zeta)= 
{{2}\over{\gamma+2}}\int_\zeta^\infty\eta^{-{{\theta}\over{\gamma+2}}}
\Phi(\eta)d\eta-{{2}\over{\gamma+2}}\zeta^{{1}\over{\gamma+2}}\int_\zeta^\infty
\eta^{-{{\theta+1}\over{\gamma+2}}}\Phi(\eta)d\eta
\end{equation}
In order to eliminate the integral we differentiate  this
equation twice with respect to $\zeta$ to reduce it to the third-order 
differential equation 
\begin{eqnarray}
\zeta^2\Phi^{\prime\prime\prime}(\zeta) &+&\zeta\Big ((3-{{2\theta+1} 
\over{\gamma+2}}) + {{\zeta}\over{2}}\Big 
)\Phi^{\prime\prime}(\zeta)  +  \Big ((1-{{\theta}\over{\gamma 
+2}})(1-{{\theta+1}\over{\gamma+2}})+ 
{{3\gamma+5-2\theta}\over{2(\gamma+2)}}\zeta\Big )\Phi^\prime(\zeta) 
\nonumber \\ & +& \Big ({{1}\over{2}}(1-{{\theta}\over{\gamma
+2}})(1-{{\theta+1}\over{\gamma+2}})-{{2}\over{(\gamma+2)^2}}\Big 
)\Phi(\zeta)=0.
\end{eqnarray}
We can rescale the equation to obtain
\begin{eqnarray}
{{\zeta^2}\over{4}}\Phi^{\prime\prime\prime}(\zeta)&+&{{\zeta}\over{2}}\Big 
((3-{{2\theta+1}\over{\gamma+2}})+{{\zeta}\over{2}}\Big) 
\Phi^{\prime\prime}(\zeta)+\Big ((1-{{\theta}\over{\gamma
+2}})(1-{{\theta+1}\over{\gamma+2}})+{{3\gamma+5-2\theta}\over{2(\gamma+ 
2)}}{{\zeta}\over{2}}\Big )\Phi^\prime(\zeta)\nonumber \\ & +& \Big 
((1-{{\theta}\over{\gamma+2}})(1-{{\theta+1}\over{\gamma+2}})- 
{{4}\over{(\gamma+2)^2}}\Big )\Phi(\zeta)=0.
\end{eqnarray}
The solution of equation (15) is given by the generalised 
hypergeometric function
\begin{equation}
\Phi(\zeta)= \ _2F_2\Big 
(1+{{\sqrt{17}-2\theta-1}\over{2(\gamma+2)}},1-{{\sqrt{17}+2\theta+1} 
\over{2(\gamma+2)}};1-{{\theta}\over{\gamma+2}},1-{{\theta+1} 
\over{\gamma+2}}; -{{\zeta}\over{2}}\Big ). 
\end{equation}
This is the explicit and exact scaling solution from which one can find 
the large-$\zeta$ behaviour,
\begin{equation}
\Phi(\zeta) \sim e^{-\zeta/2}.
\end{equation}
The two scaling functions are related through 
\begin{equation}
\phi(xt^{{{1}\over{\gamma+2}}})=(xt^{{{1}\over{\gamma+2}}})^{-\theta}\Phi 
((xt^{{{1}\over{\gamma+2}}})^{\gamma+2})).
\end{equation}
Hence we obtain the scaling function for large $\zeta=\xi^{\gamma+2}$ 
as 
\begin{equation}
\phi(\xi) \sim \xi^{-(1+a)}e^{-{{\xi^{\gamma+2}}\over{2}}}
\end{equation}
One can recover the solution  obtained in \cite{kn:has94} using 
different and indirect method from this general solution by setting 
$\gamma=0$. 
It is possible to obtain scaling solution for higher value of $\alpha$ 
but as $\alpha$ increases the numerator and denominator parameter of the 
generalised hypergeometric function becomes  increasingly 
complicated.  However, the knowledge of the $z$ and 
$\theta$ 
and details survey reveals that it is possible to write the scaling 
solution in the large $\xi$ limit for general $\alpha$ as  
\begin{equation}
\phi(\xi) \sim \xi^{-(1+a(\alpha))}e^{-{{\xi^{\alpha+\gamma+1}}\over{2}}},
\end{equation}
 where, $a(\alpha)$ is the solution of the following equation for $q$
\begin{equation}
2{{\Gamma(q+1)\Gamma(\alpha+1)}\over{\Gamma
(q+\alpha+2)}}={{1}\over{\alpha+1}}
\end{equation}
Therefore, we can write the empty size distribution for the long time 
limit as
\begin{equation}
\psi(x,t) \sim x^{-(1+a(\alpha))}\Phi(\zeta).
\end{equation}
That is, we can choose scales 
$\psi_0(x)= x^{-(1+a(\alpha))}$ depending on spatial variable for the 
empty size distribution function and $t_0(x)= x^{-(\alpha+\gamma+1)}$ for 
the temporal variable. Therefore,  in the new scale the properties of the 
empty size distribution function can be expressed 
in terms of one variable i.e.,
\begin{equation}
\psi(x) \sim x^{-(1+a(\alpha))} e^{{{t}\over{t_0}}}.
\end{equation}
It implies that $\psi/\psi_0$ and $t/t_0$ are 
self-similar coordinates.

\subsection{Statistically self-similar pattern formation}

The existence of scaling shows that the pattern created in the long time 
limit becomes scale free  i.e. the whole  can be obtained from 
the part by a suitable change in scale. Essentially, this  implies  that we 
can invoke the idea of a fractal dimension: a 
dimension that uniquely determines the geometry of the object. 
We now use the usual box counting method to determine the fractal 
dimension. We define size of the segments to be
\begin{equation}
\delta = {{M(1,t)}\over{M(0,t)}} \sim t^{-{{1}\over{\gamma+2}}},
\end{equation}
and we count the number of such segments needed to cover the whole set of 
points to determine the fractal dimension. In the limit $\delta 
\rightarrow 0$, we find that the number of segments $<N(\delta)>$ 
requires to cover the set created by the Eq. (4) scales as
\begin{equation}
<N(\delta)> \sim \delta^{-D_f(\alpha)}.
\end{equation}
Where, $D_f(\alpha)$ is the real and positive root of the polynomial
equation in $q$ obtained from the Eq. (6). Consequently, $D_f$ is found to 
obey 
\begin{equation}
2{{\Gamma\Big (D_f(\alpha)+1\Big )\Gamma(\alpha+1)}\over{\Gamma
\Big (D_f(\alpha)+\alpha+2\Big )}}= {{1}\over{\alpha+1}}.
\end{equation}
Thus, a single scaling exponent $D_f(\alpha)$ completely characterizes 
the structures of the objects which is  reminiscent of the jamming limit. 
Note that fractal dimension $D_f(\alpha)$ does not 
depend on the exponent $\gamma$ so it is independent of the rate at which 
particles are deposited, provided $\gamma > -(\alpha +1)$. In Table 2 we 
give a spectrum of fractal dimensions for different values of $\alpha$ 
with $\beta=0$. This 
shows that as $\alpha$ increases the fractal dimension decreases. Later, 
we also attempt to give a physical interpretation of the exponent $\alpha$.

\section{Model II}

We shall now consider another model that follow the same parking 
distribution (i.e. $p(z) \sim z^{\alpha-1}$), but 
different deposition rate. The deposition rate of this  model is 
\begin{equation}
F(x,y|z)=xyz
\end{equation}
This particular choice of the deposition rate implies that all the 
points along the chosen empty space are not equally likely to be 
deposited. Although all the empty spaces compete on equal footing to be 
chosen where particles can be deposited.  That is, the rate depends on the 
size of the deposited 
particles as well as on the size of the two smaller empty spaces created 
due to deposition. Substituting 
this into the equation Eq. (1) with $n=3$ and $m=2$ we obtain the 
following 
rate equation 
\begin{equation} 
{{\partial \psi(x,t)}\over{\partial 
t}} = -{{ x^{\alpha+4}}\over{(\alpha+3)(\alpha+4)}}\psi(x,t) 
+2 \int_x^l x(y-x)^{\alpha+2}\psi(y,t)dy \hspace{4mm} {\rm for} 
\hspace{4mm} x\leq l. 
\end{equation}
Notice that $\alpha=-1$ describe the fragmentation process with the 
fragmentation kernel $F(x,y)=xy$ \cite{kn:has4}. Also notice that the 
intensity of non-local character is higher than 
the previous model for which we find it is increasingly difficult to 
find scaling solution. Nevertheless, for our purpose it is enough to 
know that scaling exist. Substituting  
the definition of the moment into the above equation yields, 
\begin{equation}
{{dM(n,t)}\over{dt}}=\Big [ {{\Gamma(n+2)\Gamma(\alpha 
+5)}\over{\Gamma(n+\alpha+5)}}-1 \Big ]M(n+\alpha +4,t).
\end{equation}
We find that the asymptotic behaviour of the moment can 
provide some of the interesting features of the system. Hence,
from now on we are only interested in finding the fractal dimension 
of the system that can uniquely characterize the structure. Following a 
similar procedure as we have done before we find that  
the number of segments needed to cover the whole set created, scales as
\begin{equation}
<N(\delta)> \sim \delta^{-D_f(\alpha)},
\end{equation}
where, $D_f(\alpha)$ as before can be obtained to satisfy
\begin{equation}
{{\Gamma\Big (D_f(\alpha)+2\Big )\Gamma(\alpha
+5)}\over{\Gamma\Big (D_f(\alpha)+\alpha+5\Big )}}=1.
\end{equation}
Here, the $D_f(\alpha)$ is  the fractal dimension of the
pattern formed by this model. It is clear that the order of the  
polynomial equation is determined by the 
$\alpha$ value. Therefore the fractal dimension $D_f(\alpha)$ is 
different for different $\alpha$ values. Comparing this 
model with the previous model we find that fractal dimension for this 
model is always higher than that for the previous model for each 
corresponding  $\alpha$ value. Also, in both cases fractal dimensions 
appear to decrease 
monotonically as $\alpha$ increases. This is a feature that we shall discuss 
further later. We  intend to determine if there exists any relation 
between the  degree of order in  the pattern and the fractal dimension. 
In order to  do this we consider a further generalization of the two 
models we discussed.

\section{Model III}

\vspace{6mm}

We now turn to the more general model in which, at each time step, more than 
one particle will attempt to be deposited. We choose the parking 
distribution  and deposition rates which are of the same functional form 
i.e., we choose the particle size distribution for deposition to take the 
form 
\begin{equation} p(x_i)=g(x_i) \hspace{4mm} {\rm for} \hspace{4mm} x\leq l,
\end{equation}
and the deposition rate
\begin{equation}
F(x_{n-m+1},..,x_n|x_1,..,x_{n-m}) =\prod_{i=n-m+1}^ng(x_i).
\end{equation}
Substituting (29) and (30) into the equation (1) we obtain the 
following rate equation
\begin{equation}
{{\partial\psi(x,t)}\over{\partial
t}}=-F_n(x)\psi(x,t)+mg(x)\int_x^l\psi(y,t)F_{n-1}(y-x)dy \hspace{4mm} 
{\rm for} \hspace{4mm} x\leq l, 
\end{equation}
where, the functions  $\{F_n(x), n=2,3...\}$ are defined by
\begin{equation}
 F_n(x)= \int\delta(x-\sum_{i=1}^nx_i)\prod g(x_i)dx_i.
\end{equation}
Eq. (35) is equivalent to the dynamic system of breaking an interval 
into $n$ pieces and throwing away $(n-m)$ of them at each time step 
\cite{kn:has35}. It is straightforward to 
show that \begin{equation} F_n(x)=\int_0^xg(y)F_{n-1}(x-y)dy,
\end{equation}
for $n \geq 3$ and
\begin{equation}
F_2(x)=\int_0^xg(y)(x-y)dy.
\end{equation}
We further specify our model by choosing $g(y)=y^\beta$, with $\beta$ 
treated as an external tunable parameter.  When 
$g(y)=y^\beta$ we can obtain 
the function $\{F_n(x),n=2,3,..\}$ from equations (32) and (33) as 

\begin{equation}
F_n(x)={{[\Gamma(\beta+1)]^n}\over{\Gamma(n(\beta+1))}} x^{n(\beta+1)-1}. 
\end{equation}
It is interesting to note that if we set $\beta=0$, $m=2$ and 
$n=\alpha+2$ in equation (35) we get the same rate equation for the 
empty size distribution as described by the model I (equation (4)) with 
$\gamma=0$.   It reveals that we can give an alternative 
interpretation of the model I.  That is, at each time step an interval is 
broken  into $\alpha+2$ random pieces  and $\alpha$ of them removed from the 
system.  Alternatively, we can say that the exponent $\alpha$ determines 
the number of  particles to be deposited at each time step on an interval. 
However, for $\alpha>1$ the 
$m$ value put a strong constraints on the depositing particles. That is, 
the process describe sequential deposition of $\alpha$ particles 
consecutively as if they were a single particle.  
Similarly, if we  set $\beta=1$, $m=2$ and $n={{\alpha+5}\over{2}}$ we 
get the same rate  equation as described by the model II. This reveals 
that at each time step ${{\alpha+1}\over{2}}$ are deposited 
consecutively. These two feature helps us to understand the physical role 
played by the parking distribution exponent $\alpha$. That is, as 
$\alpha$ increases the length of the depositing particles on the average 
increases with respect to that of corresponding lower $\alpha$ value. In 
order for further support we concentrate on the fractal dimension of the 
resulting pattern.

Since the moments of the empty space distribution 
function can cheracterise the fragmenting systems more easily than the 
empty spaces distribution function itself, we now 
consider the behaviour of the moment only. For $g(y)=y^\beta$, 
the time evolution of the moments can be obtained using (5) 
\begin{equation}
{{dM(q,t)}\over{dt}}=[\Gamma(\beta+1)]^{n-1}\Big ({{m\Gamma(q+\beta+1)}\over
{\Gamma(q+n(\beta+1))}}-{{\Gamma(\beta+1)}\over{\Gamma(n(\beta+1))}}\Big )
M(q+n(\beta+1)-1,t).
\end{equation}
This equation can be solved  for the moments $M(q,t)$, the 
solution again taking the form  of a generalised hypergeometric 
function with $(n-1)(\beta+1)$ numerator and denominator 
parameters.
 We now consider the scaling behaviour of these models
that define the scaling exponent $\theta$ and $z$, and give the 
long time dependence of the moment as $M(q,t) \sim t^{z(\theta-q-1)}$. 
We can immediately find $z$ for all $m$, $n$ and $\alpha$, because, in the 
long time limit, the moments behave as
\begin{equation} M(q,t) \sim 
A(q)t^{-b(q)} \end{equation}
Substituting this into the rate equation for the moment yields a 
difference equation
\begin{equation}
b(q+n(\beta+1)-1)=b(q)+1.
\end{equation}
Iterating this and using the $q$ value for which the moment becomes 
time independent, we find 
\begin{equation}
b(q)={{q-q^*}\over{n(\beta+1)-1}}.
\end{equation}
This gives 
\begin{equation}
z={{1}\over{n(\beta+1)-1}},
\end{equation}
and $q(\beta,m,n)$ related to  $ \theta$ by $\theta=D_f+1$ and can be 
obtained from (36) and satisfies
\begin{equation}
{{m\Gamma(q+\beta+1)}\over
{\Gamma(q+n(\beta+1))}}={{\Gamma(\beta+1)}\over{\Gamma(n(\beta+1))}}.
\end{equation}
In particular, this model can be described as a correlated sequential 
deposition of particles on a substrate, where, the degree of correlation 
is determined by the exponent $\beta$. 

For power-law form of parking distribution function this model describes 
the deposition of $(n-m)=1,2,3,...$ etc. particles 
and creating $m=2,3,...$etc. number of new empty spaces respectively. If 
$(n-m)=p$ and $m<(p+1)$ then the  $m$ value  determines how many of them 
deposit consecutively. This deposition phenomenon  can 
equivalently be interpreted as cutting an interval into $n$ pieces and 
removing  $(n-m)$ of the parts created and  repeating the process with the 
remaining $m$  pieces thus resembling the concept of classic Cantor set. 
The fact that  the size and position of the particles to be removed are 
chosen  stochastically, unlike in the Cantor set where there are no 
kinetics, is  irrelevant because we are considering the scaling regime, 
$t\rightarrow  \infty$.  Since this process is repeated {\it ad 
infinitum}, it  forms  stochastic fractals with dimension between 
$0\leq  D_f\leq 1$. As before, we use the box counting method by defining a  
characteristic  length $\delta$ so that we can count the number of 
segments required  to cover the set when $\delta\rightarrow 0$ which 
determines the properties of the resulting set and scales as, 
\begin{equation}
<N(\delta)> \sim \delta^{-D_f}. 
\end{equation}
Thus we find that the  Hausdorff-Besicovitch dimension in this case is 
equal to $q^*$. We still have $z$ given by equation (41), but now the 
value of $q^*$ (and hence $\theta$ and $D_f$) are non-trivial. 
This class of fractals includes that considered in 
\cite{kn:krapivsky}, with $n=3$, $m=2$ and $\beta=0$.
In the limit $\beta \rightarrow \infty$ we can use Stirling's 
formula  and the following asymptotic formula
\begin{equation}
\Gamma(az+b) \sim \sqrt{2\pi}e^{-az}(az)^{az+b-{{1}\over{2}}},
\end{equation}
in equation (42) to find that $q^* (=D_f)\rightarrow 
\ln(m)/\ln(n)$. This coincides with the fractal dimension of the classic 
Cantor set.  That is in the limit 
$\beta \rightarrow \infty$ the standard deviation of the size of the 
pieces created tends to zero. This is easily verified by calculating the 
mean and standard deviation from $F\{x_i\}$. This particular finding 
implies that in this limit $F_n(x)$ behaves approximately as
\begin{equation}
F_n(x)=x^\lambda\int \delta(x_1-x_3)dx_1\int \delta(x_1-x_2)dx_2...\int 
\delta(x_1-(x-\sum_{i=1}^{n-1}x_i))dx_{n-1}.
\end{equation}
Substituting this into  equation (32) we obtain
\begin{equation}
{{\partial \psi(x,t)}\over{\partial 
t}}=-{{x^\lambda}\over{n}}\psi(x,t)+m(nx)^\lambda\psi(nx,t).
\end{equation}
This describes a model that splits an interval into $n$ equal pieces and 
keeps only  $m$ of them. Substituting the definition of the moment Eq. 
(5) into Eq. (46) we obtain
\begin{equation}
{{dM(q,t)}\over{dt}}=\Big [{{m}\over{n^{q+1}}}-{{1}\over{n}}\Big 
]M(q+\lambda,t).
\end{equation}
The solution of the Eq. (4.44) is
\begin{equation}
M(q,t) \sim t^{-{{q-D_f}\over{\lambda}}},
\end{equation}
and immediately reveals that $D_f={{\ln m}\over{\ln n}}$ as in the classic 
Cantor set with kinetic exponent ${{1}\over{\lambda}}$.
In the limit $\beta +1 \rightarrow 0$ we can analyse Eq. (42) to show 
that $q^*$ (and $D_f$) tends to zero 
like $\gamma(m,n)(\beta+1)$ 
where $\gamma(m,n)=n(m-1)/(n-m)$. Consequently, we see that for 
all $m$ and $n$ there is a spectrum of fractal dimensions 
between $\beta\rightarrow -1$ when $D_f \rightarrow 0$ and 
$\beta \rightarrow \infty$ when $D_f \rightarrow 
\ln(m)/\ln(n)$. This is a very striking result. It implies that in the 
limit $\beta \rightarrow \infty$ particles are only deposited in the 
center of the empty space and produce strictly self-similar patterns. 

\section{Summary and discussion}

In this work we presented a number interesting results relating to 
random and correlated sequential deposition of a mixture of particles of 
finite sizes. We find that the pattern created is statistically scale 
invariant. We also attempt to show the relationship between the fractal 
dimension and the degree of order in the resulting pattern created in 
the long time limit. In table 2 we present some values of the fractal 
dimension 
as a function of $m$ and $n$ for $\beta=0$ using model III. The 
corresponding fractal 
dimension for the Cantor set are given in parentheses. In table 1 we give 
the fractal dimension for $m=2$ and $n=3$ for some different 
values of $\beta$. Table 1, and a more numerical survey 
confirm that the fractal dimension increases monotonically as
$\beta$ increases. Moreover, fractal dimension  appears to  decrease as $n$ 
increases for  a given $m$ and vice versa, provided 
$(n-m)>0$. In table 3, we compare the fractal 
dimensions for   different values of $\alpha$ for model I and II. This 
table and further detail numerical 
survey confirms that for the same $\alpha$ value, the model II creates 
pattern that has higher fractal dimension than that for model I. We 
further notice that  the first row of table 2 for fractal dimension with 
$m=2$ are exactly the same as in table 3 with $\beta=0$ and $\alpha=1,2,3$. 
Similarly, the fractal dimension obtained from model III for 
$\beta=1$, $m=2$ and $n={{\alpha+5}\over{2}}$ would be the same as 
obtained from model II for corresponding 
${{\alpha+1}\over{2}}=1,2,...,etc.$. These, 
further shows that as $\alpha$ increases the length of the deposited 
particles on the average increases with respect to that of a 
corresponding average length
for a lower $\alpha$ value. Which appear to be consistent with our details 
survey that reveals that 
fractal dimension monotonically decreases as $\alpha$ increases and as 
$\alpha \rightarrow \infty$ the fractal dimension $D_f\rightarrow 0$.  
These observations immediately confirms that the exponent $\alpha$ does 
not play any role  in creating ordered pattern since fractal 
dimension does not reach to constant non-zero value as $\alpha 
\rightarrow \infty$.

In a recent letter \cite{kn:brilliantov}, Brilliantov {\it et. al} 
studied the random sequential adsorption of a mixture of particles with 
a continuous distribution of sizes determined by the power-law form 
(equation (2)). 
They reported that  the pattern created in the long time becomes more and 
more ordered as 
$\alpha$ increases and in two dimensions it reaches the Apollonian 
packing in the limit $\alpha \rightarrow \infty$ when the depositing 
particles are finite mixture of disk. It is important to notice that the 
one dimensional analogue of this model is deposition of a mixture of 
rods. Evidently, one expect the fractal dimension to coincide with the 
classic Cantor set when the pattern becomes perfectly ordered pattern. In 
Ref. \cite{kn:brilliantov} an exact expression for fractal dimension also 
given for general $d$ dimension. It is also clear that 
the Apollonian packing is a non-random or 
strictly self-similar fractal like Sierpinsky gasket difference lies only 
in the geometry of the depositing particles. The observation 
that the pattern becomes more and more ordered should be true for any 
dimension for any geometry including $d=1$. 
That is, in order to support the result that in the limit $\alpha 
\rightarrow \infty$ the pattern becomes more and more ordered the fractal 
dimension must reach to constant non-zero value in that limit. In 
particular, as 
$\alpha \rightarrow \infty$ the fractal dimension must reach to the value 
$\ln 2/\ln3$ which is 
one dimensional analogue of Apollonian packing and Sierpinsky gasket both. 
In one dimension we can solve the 
model exactly which corresponds to our model I. The exact enumeration of 
fractal dimension and details numerical survey reveals that $D_f 
\rightarrow 0$ as $\alpha \rightarrow \infty$ instead of reaching a 
constant non-zero value. Therefore, it contradicts 
the analysis we  give in this work with result reported in 
\cite{kn:brilliantov}. More recently, deposition of a finite mixture 
of rectangles  in two dimensional substrate is studied in Ref. 
\cite{kn:has34}.  
Although, in this work particles are only allowed to deposit one of the 
four corner, yet, it retains the generic feature of deposition phenomena 
of a definite mixture and in particular it is very close to the model we  
consider in this work. The work 
in ref. \cite{kn:has34} is stochastic counterpart of the deterministic 
Sierpinsky carpet
or Cantor gasket \cite{kn:has91} which is strictly  self-similar. In this 
deterministic case, the initiator is a square
and the generator subdivides at each step into $b^2$  equal square of
which $p$ of them are removed according to a fixed rule. After an
infinite number of iterations the resulting set
can be seen as a  generalization of  the Cantor set to two dimensions that
has the fractal dimension $D_f=\ln (b^2-p)/\ln b$. However, in the case 
of its stochastic counterpart we have shown that the system does not 
reach to 
simple scaling instead the system shows multiscaling. A result which we 
believe to be true for 
deposition of a finite mixture particles of any geometry in more 
than one dimension. That is, the  pattern created in the long time limit has 
global scaling exponent  $D_f$ and local scaling exponent known as $f(q)$, 
where $q$ is the Holder exponent. That is, 
pattern can be divided into  a subset that scales with different fractal 
dimension, a phenomena called multifractality. However, in the case of 
strictly self-similar pattern the 
system reaches to simple scaling behaviour. The study further revealed that 
the global exponent or the 
fractal dimension of random fractal is always lower  than its 
corresponding strictly self-similar counter part. Therefore, we 
conclude that fractal dimension must increases with increasing order and 
reaches to maximum value when the pattern is in perfect order.
In this work, we show that 
the exponent $\alpha$ 
does not play any role in creating an ordered pattern. 
Instead, it implies that the length  of the deposited particle at each 
time step increases on  average  as $\alpha$ value increases. 

In order to create an ordered pattern we reveal that one needs to choose 
$F(x,y|z) \sim (xy)^\beta$ and $p(z) \sim z^\beta$. This model for 
$\beta \neq 0$ describes that at each event an empty space is chosen 
randomly. However, once 
this decision has been made, particles are more likely to deposit at 
the center of the empty space  than on either side of it. Of course the 
tendency to deposit at  the center increases  as $\beta$ increases. In fact, 
our analysis 
further supports that the fractal  dimension increases with the degree of 
increasing order and reaches its 
maximum value (a non-zero constant) in the perfectly ordered pattern, as it 
does in the classic 
Cantor set or in the Seirpinsky gasket. Krapivsky {\it et al.} reported 
in \cite{kn:krapivsky} that the dimension of the random fractal is 
always smaller than its deterministic counterpart. If we look at the 
situation for  the stochastic Cantor set, we find that in order to get 
the deterministic  classic set $\Big ({{\ln m}\over{\ln n}}\Big )$, 
$\alpha$ alone does not play any role in  creating any ordered pattern.
Instead, one has to choose $F(x,y|z)$ to be the same power law form as for 
parking distribution $p(z)$. In this case only the $\alpha$ value 
determines the degree of tendency to place the particles in the center of 
the empty space and in the limit $\alpha\rightarrow \infty$ particles are 
always placed exactly at the center of the empty space. We can 
quantify the increasing regularity of the resulting pattern 
created in the long time limit by introducing the concept of 
entropy production that characterizes the degree of order as
\begin{equation}
S=-\sum_{C_k}p(C_k)\log p(C_k). 
\end{equation}
Since in the $\beta \rightarrow \infty$ limit there is only one definite 
configuration we have $ p(C_k)=1$ that contributes to the entropy. In 
fact we can define
\begin{equation}
p(C_k)={{D_f}\over{\ln m/\ln n}},
\end{equation}
where, each $D_f$ corresponds to one definite configuration. We find that 
$S$ increases as $\beta$ decreases towards zero. That is, as $\beta$ 
decreases the  number of possible  configurations increases.  In this 
work, we 
generalize the conventional RSA where the position of a particle to be 
deposited in the empty space is chosen randomly and that the degree of 
randomness by the position dependent deposition rate. That is, the position 
where the particle is to be placed  is 
chosen by the size of the empty space being destroyed and by the size of 
the empty spaces created on either side. As a prospect of 
future work one can choose $p(z)=\delta(x-1)$ and 
$F(x,y|z)$ to be position 
dependent. Studying  RSA with these choices the system obviously 
will reach a jamming limit, but how  it varies with the degree of order
can be of greater physical interest.

\newpage
\begin{center}
\begin{tabular}{|l|l|} \hline
 $\beta$ & $ D_f$ \\
\hline
$-{{1}\over{2}}$ & ${{1}\over{2}}$ \\
0 & 0.5616288 \\
${{1}\over{2}}$ & 0.5841 \\
1 & 0.5956 \\
2 & 0.6073 \\
$\infty$ & 0.6309 \\
\hline
\end{tabular}
\end{center}
Table 1: The fractal dimension $D_f$ of stochastic fractals for $m=2$ and 
$n=3$ for increasing $\beta$ values. 

\vspace{3mm}

\begin{center}
\begin{tabular}{|l|l|l|l|}\hline
$m$ & $n$ \\ \cline{2-4}
& $3$ & $4$ & $5$ \\ \hline
$2$ & $0.5615288 (0.6309)$ & $0.4348 (1/2)$ &$0.3723(0.4307)$ \\
\hline
$3$ &  $1$ & $0.7478 (0.7925)$ & $0.6295 (0.6826)$ \\ \hline
$4$ & &$1$ & $0.8315(0.8614)$ \\ \hline
$5$ & & & $1$ \\ \hline
\hline
\end{tabular}
\end{center}
Table 2: The fractal dimension $D_f$ of the stochastic fractals
with $\beta=0$; the corresponding dimensionality for the
Cantor set ($\beta=\infty$) is given in the parentheses.

\vspace{3mm}

\begin{center}
\begin{tabular}{|l|l|l|}\hline
$\alpha$ & $\beta=0$ & $\beta=1$ \\ \cline{1-3}
$1$ & $0.5616$ & $0.5956$ \\  \hline
$2$ & $0.4348$ & $0.5183$   \\  \hline
$3$ & $0.3723$ & $0.466542$   \\ \hline
$4$ & $0.33405$ & $0.429121$   \\  \hline
$5$ & $0.30784$ & $0.400614$  \\ \hline
$6$ & $0.288505$ & $0.37805$  \\ \hline
$7$ & $0.27351$  & $0.35966$  \\ \hline
\hline
\end{tabular}
\end{center}
Table 3: The fractal Dimension $D_f$ for $\beta =0,1$ for different
values of $\alpha$

\vspace {8mm}

{\large \bf Acknowledgement}

\vspace {8mm}

The author is indebted to G. J. Rodgers for numerous discussion and valuable 
remarks during this work.

\end{document}